\documentclass[conference]{IEEEtran}
\usepackage{mathptmx}  \usepackage{amsmath}
\usepackage{amsfonts}
\usepackage{amssymb}
\usepackage{xspace}
\usepackage{graphicx}
\usepackage{psfrag}
\usepackage{pifont}
\usepackage{color}    \usepackage{geometry}
\usepackage{subfigure}
\usepackage[hyphens]{url}

\def\centerhack#1{\hbox to 0pt{\hss\footnotesize #1\hss}}

\renewcommand{\baselinestretch}{0.98}
\makeatletter
\def\@listi{\leftmargin\leftmargini
    \parsep 1\p@ \@plus0\p@ \@minus\p@
    \topsep 2\p@   \@plus0\p@ \@minus\p@
    \itemsep1\p@ \@plus0\p@ \@minus\p@}
\let\@listI\@listi\@listi
\makeatother

\newcommand{\etal}{et~al.\@\xspace} 
\newcommand{\eg}{e.g.,\xspace}
\newcommand{\ie}{i.e.,\xspace}

\newcommand{\chen}[1]{}

\newcommand{\scion}{\xspace{}SCION\xspace}
\newcommand{\ccn}{\xspace{}CCN\xspace}
\newcommand{\ndn}{\xspace{}NDN\xspace}
\newcommand{\nebula}{\xspace{}NEBULA\xspace}
\newcommand{\ip}{\xspace{}IP\xspace}

\def\centerhack#1{\hbox to 0pt{\hss\footnotesize #1\hss}}
\def\dchack#1{\vbox to 0pt{\vss{\hbox to 0pt{\hss#1\hss}}\vss}}

\newcommand{\xmark}{\ding{55}}
\newcommand{\cmark}{\ding{51}}

\title{Modeling Data-Plane Power Consumption of \\ Future 
Internet Architectures}

\author{\IEEEauthorblockN{Chen Chen}
	\IEEEauthorblockA{ETH Zurich}
	\and
	\IEEEauthorblockN{David Barrera}
	\IEEEauthorblockA{ETH Zurich}
	\and
	\IEEEauthorblockN{Adrian Perrig}
	\IEEEauthorblockA{ETH Zurich}
	}

\begin{document}

\maketitle

\begin{abstract}

With current efforts to design Future Internet Architectures (FIAs), the 
evaluation and comparison of different proposals is an interesting research challenge. 
Previously, metrics such as bandwidth or latency have commonly been used 
to 
compare FIAs to IP networks. We suggest the use of \textit{power consumption} as 
a metric to compare FIAs. While low power consumption is an important goal in 
its own right (as lower energy use translates to smaller environmental impact 
as well as lower operating costs), power consumption can also serve as a 
proxy for other metrics such as bandwidth and processor load. 

Lacking power consumption statistics about either commodity FIA routers or 
widely deployed FIA testbeds, we propose models for power consumption of FIA 
routers. Based on our models, we simulate scenarios 
for measuring power consumption of content delivery in different FIAs. 
Specifically, we address two questions: 1) which of the 
proposed FIA candidates achieves the lowest energy footprint; and 2) 
which set of design choices yields a power-efficient network architecture?
 Although the lack of real-world data makes numerous 
assumptions necessary for our analysis, we explore the uncertainty of our 
calculations through sensitivity analysis of input parameters.

 \end{abstract}

\section{Introduction}
\label{sec:introduction}\label{sec:intro}

The current Internet requires a considerable amount of power, 
consuming nearly 1\% of annual electricity production 
worldwide~\cite{hinton2011power}. Around 
50GW of power is consumed by network equipment, 
and this number is expected to double by 2020~\cite{vereecken2008estimating}. 
Increased power consumption not only implies greater monetary cost, 
but also exerts an expanding environmental impact such as carbon footprint~\cite{berkeley_energy_impact, shehabi2014energy} and 
pollution~\cite{power_pollution_internet}. Reversing the trend is both 
imperative and rewarding. In fact, a 10\% reduction of global network power 
consumption could eliminate the need for 5 recent large nuclear reactors today~\cite{nuclear_reactor}.

The past five years have seen significant effort from the research community to 
design Future Internet Architectures (FIAs).  The underlying goals and designs of 
FIAs can vary drastically. For example, Named Data 
Networking (\ndn~\cite{Jacobson2009NDN}) treats 
content (rather than end hosts) as the principal entity and enables efficient 
content distribution. Mobility 
First~\cite{raychaudhuri2012mobilityfirst} treats mobile devices as 
first-class citizens on the network. 
\nebula~\cite{Anderson2013NEBULA} provisions a highly-available 
and extensible core network interconnecting data centers. 
XIA~\cite{anand2011xia} enables evolution of the network's underlying protocol stacks.
\scion~\cite{Xin2011SCION} enables highly-available communication.
With such diverse design goals, evaluating and comparing FIAs under a common 
framework is a difficult task. 

Although metrics such as bandwidth or latency have 
been used to compare the IP network to FIAs~\cite{fayazbakhsh2013less}, and 
power consumption of Content Centric Network (\ccn) has been evaluated~\cite{perino2011reality}, 
to our knowledge power consumption as a metric has not yet 
been considered for comparison of different FIAs. 
Consequently, answers to key design questions such as ``does fetching content 
directly from a remote server require less power than 
retrieving the same content from a nearby cache?'' and
``which is more power 
efficient: packet-carried state or routing table lookups?''
 remain largely unknown
and yield some counter-intuitive answers as we show in this paper.

In certain cases, the power implication of some FIA design choices is 
straightforward. For example, some 
architectures require routers to 
conduct cryptographic operations for security reasons~\cite{Anderson2013NEBULA, 
Xin2011SCION}, which inevitably increases router computation, and thus increases 
router power consumption. However, the power consumption implications of other 
FIA design choices are difficult to pre-determine because the design choices 
introduce trade-offs. For example, 
architectures
that use Packet-Carried State (PCS) do not require routing tables, which 
reduces the routers' power consumption. However, 
using PCS requires embedding extra forwarding information 
in packet headers, which 
increases the number of bits that must be transmitted, and therefore increases 
the power consumed to forward the packets.
 
As a first step towards analyzing the power consumption of the IP network and FIAs, 
we focus on the power consumption of the data-plane. Since data-plane traffic 
consumes 83\% of the total power consumed by 
the Internet (compared to 17\% consumed by 
control-plane~\cite{bolla2011energy}), we believe our analysis covers the 
largest component of power consumption in IP networks and FIAs.

Three main challenges exist in analyzing power consumption of FIAs. First, 
modeling FIA router forwarding behavior is more complex than modeling today's IP 
routers which are themselves non-trivial to model. Levels of abstraction for 
router models vary, ranging from gate-level 
modeling~\cite{wang2002power}, to microarchitecture-level 
modeling~\cite{peh2001delay}, to router-level modeling~\cite{hinton2011power}. 
Choosing the correct level of abstraction is necessary to preserve the model's 
simplicity, yet still make it useful to highlight the differences of various FIA 
designs.

To address the first challenge, we present a generic FIA router model 
that captures the commonality of IP and FIA routers as well as the 
peculiarities of FIA routers (Section~\ref{sec:model}). We use a hybrid method 
that combines two models at different levels: 1) a high-level model to 
characterize 
the power consumed by the common behaviors shared between IP routers and FIA 
routers; and 2) a low-level model to characterize the power consumed by the 
behaviors unique to FIA routers.

The second challenge is that power consumption analysis of network data-planes 
requires analysis of router forwarding behaviors in the context of real-world
network topologies and workloads. For example, \ndn~\cite{Jacobson2009NDN} 
requires routers to host content caches, which increase routers' power 
consumption. However, content caches reduce the number of network links the 
queries and responses traverse. Thus, the overall power consumed to process 
these queries may be reduced. Whether the reduced power can compensate for the 
power expended to operate the content cache largely depends on how routers 
inter-connect and the temporal locality of the workload.

To address the second challenge, we conduct a large-scale 
simulation of content delivery traffic across multiple autonomous systems 
(ASes) to analyze power 
consumption of different network architectures (Section~\ref{sec:simulation}). 
Particularly, we focus on using the simulation to investigate the influence of 
content caches and packet-carried forwarding information on power 
consumption.

The third challenge is to define a common comparison framework for different FIAs. 
The framework should provide useful clues about power-efficient FIA designs, 
independent of specific design choices of particular architectures (\eg the 
hash function used). 
For our comparison framework, we generalize common FIA design choices. We 
evaluate the influence of packet forwarding techniques and cache placement 
strategies on power consumption of individual routers 
and on the entire network.

In this paper, we make the following contributions: 
\begin{itemize}
\item We present the first work comparing the IP network and FIAs using power 
consumption as a metric. Our comparison framework allows us to identify 
power-efficient FIAs as well as guide designs of power-efficient FIAs.
\item We propose a generic model to characterize the forwarding behaviors of 
FIA routers and conduct a large-scale simulation based on our router model 
to analyze the power consumption of network architectures.
\item We evaluate the influence of two architectural design choices 
(related to packet forwarding and cache placement) on power 
consumption, and find that packet-carried state is generally more power 
efficient. We also find that caching, while advantageous in reducing latency, 
does not offer substantial reductions in power consumption. 
\end{itemize}

\section{Background}
\label{sec:background}\label{sec:bckgrnd}
In this section, we describe two architectural dimensions that enable 
different design choices that are common to many FIA 
designs: 1) forwarding technique, i.e., making forwarding decisions by 
routing table lookup versus packet-carried state; 2) cache placement, 
i.e., caching content by pervasive caching versus edge caching. 

\subsection{Forwarding technique: Routing Table\\ Lookup versus 
Packet-carried State}
To route a packet through the Internet, routing state
can be kept either in routing tables constructed and maintained by 
individual routers or carried in packets themselves. In the latter case, packet 
headers contain information about paths that these packets traverse. We denote 
these two methods of making forwarding decisions as Routing Table Lookup (RTL) 
and Packet-Carried State (PCS), respectively. 

RTLs relieve hosts from tasks such as path management and keep 
packet headers small. In fact, RTLs prevail in intra- and inter-domain 
routing protocols today. However, RTL consumes considerable amounts of power. 
BGP routers maintain and search routing tables containing more than 500K
entries~\cite{512k_outages}. Large routing tables 
mandate large RAM size to store the tables. In order to match the forwarding 
speed with increasing link speeds, Application-Specific Integrated Circuits 
(ASIC) with Ternary Content-Addressable Memory (TCAM) are installed on line 
cards. TCAM chips are expensive and power-hungry.
In fact, routing table lookups consume about 32\% of the entire power 
consumption of IP routers~\cite{bolla2011energy}.

Among FIAs, \ndn adopts RTL to make forwarding decisions. The \ndn routers
use Forwarding Information Base (FIB) to route users'
interests for contents. FIB conducts longest-prefix match using content names, 
instead of IP addresses. We expect \ndn FIB to be much larger than routing tables
in IP routers given the number of possible content names. Perino and Varvello~\cite{perino2011reality} estimate that a \ndn FIB contains 
up to 20M records.

In comparison, some FIA designs adopt PCS for making forwarding decisions.
 PCS allows end hosts to control the paths that packets traverse. 
PCS simplifies and speeds up packet forwarding on routers, since 
searching for matches in routing tables is not required. 
An open question is whether PCS helps reduce 
overall power consumption. PCS reduces routers' power consumption by 
removing the need for expensive routing table lookups, but PCS enlarges the size of each 
packet (by adding path information in packet headers), and thus requires 
additional power to transmit the extra bits. Moreover, FIAs like \scion and 
\nebula mandate cryptographic operations when processing the state carried 
in packets, which also adds to the power consumed by routers.

\nebula and \scion, adopt PCS to forward packets. We assume that \nebula uses
ICING~\cite{naous2011verifying} as its data plane. In ICING,  
packet headers contain Proof-of-Consents (PoCs), which certify the providers' 
consent to carry the packets, and Proof-of-Provence (PoPs), which allow upstream
nodes to prove to the downstream nodes that the upstream nodes indeed transmit the
packets. In \scion, packet headers carry a chain of Hop Fields (HFs). HFs 
carry the border routers' decisions for routing packets, but HFs are only 
meaningful to the routers that generate them. In both \nebula and \scion, 
processing the packets requires routers to process symmetric cryptographic operations and
compute Message Authentication Codes (MACs).

\subsection{Cache Placement: Edge Caching versus Pervasive Caching}

Caching content closer to the consumer to reduce network latency and bandwidth 
cost is a common practice today~\cite{Akamai}. 
This type of content caching is usually 
organized as a dedicated network of content servers, each of which resides in 
the edge network to serve local consumers' content requests. DNS redirection is 
leveraged to re-direct content requests to nearby content servers. We refer to
 this type of content caching as \textit{edge caching}.

The research community has proposed to install content caches 
directly on routers, which provides additional opportunities to further reduce 
latency and bandwidth overhead~\cite{Jacobson2009NDN}. Upon receiving a content 
request, a router can immediately reply with content if the content is 
cached locally. If the queried content cannot be served 
locally, the router can forward the request towards a different cache through 
some routing protocol. We refer to this type of content caching 
as \textit{pervasive caching} since the 
content cache could exist in both core and edge networks.

Among FIAs, \ndn proposes pervasive caching as one of its fundamental design 
principles. Each \ndn router includes a \textit{content store} which caches 
and serves content. Upon cache misses, \ndn 
forwards the packets to a nearby router that may cache the content,
and thus can reduce the length of the content delivery path.

Although edge caching and pervasive caching have been compared using
latency and bandwidth as metrics~\cite{fayazbakhsh2013less}, the difference in 
power consumption of edge caching and pervasive 
caching has not yet been explored. Pervasive caching reduces the average 
length of the paths that content queries and replies traverse, and thus
 reduces the power consumed to transmit the packets. However, 
this caching mode requires routers to host content caches. In order to match 
the speed of the line cards, routers need additional processing to rapidly 
search the existing cached content. Additional storage coupled 
with additional processing will inevitably consume more power. We explore 
these issues further in Sections~\ref{sec:model} and \ref{sec:simulation}.

\section{Modeling Power Consumption}
\label{sec:model}
In this section, we first introduce our general power consumption model  
(Section~\ref{sec:sub:overview}). Next, we use a high-level model to 
characterize the baseline power consumption of a router 
(Section~\ref{sec:sub_base}). Then we present models for the 
forwarding-decision-making module (Section~\ref{sec:sub_rt}) and content 
caching module (Section~\ref{sec:sub_cc}), to 
capture the influence of FIA design choices on power consumption. 

\subsection{Overview}
\label{sec:sub:overview}
\begin{figure}[htbp]
\centering
\includegraphics[scale=0.6]{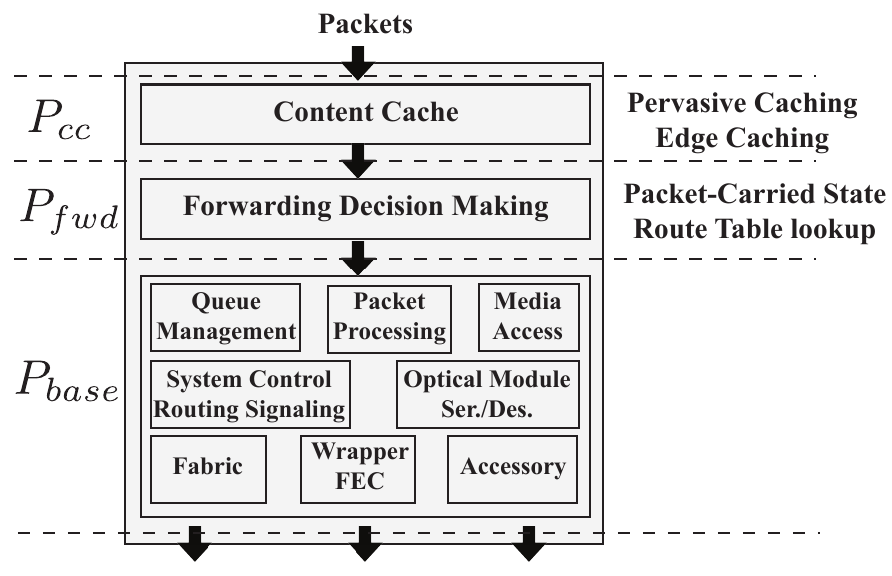}
\caption{Abstraction of the forwarding behavior of a FIA router. We present a 
similar router-component dissection as Tamm et al.~\cite{tamm2010eco}. In the 
figure, ``FEC'' is short for Forward Error Correction, ``Ser./Des.'' is short 
for Serialization and Deserialization modules and ``Accessory'' includes fans, 
power supplies, shelves, step-up converter, etc.}
\label{fig:general-model}
\end{figure}

To model power consumption, we propose a generic router model that captures 
the forwarding behaviors of both IP routers and FIA routers, as
Figure~\ref{fig:general-model} shows.
According to the two design approaches (forwarding method and caching method) 
that we investigate, we separate the content cache module and the 
forwarding-decision-making module from other router components. 
Table~\ref{tab:model-FIA} summarizes the design choices that FIAs 
use for these two different modules. 

\begin{table}[!htbp]
\centering
\begin{tabular}[!htbp]{cccc}
\hline
& \multicolumn{2}{c}{Forwarding Type} & Cache Type \\\hline
Architecture & PCS & RTL & \\\hline
\ip & & TCAM & Edge \\
\ndn & & SRAM-BF & Pervasive \\
\nebula & PoC\& PoP & & Edge \\
\scion & Hop Field & & Edge \\\hline
\end{tabular}
\caption{Methods used by network architectures for making forwarding decisions 
and caching content.}
\label{tab:model-FIA}
\end{table}

We group the rest of the router components, such as queue management and 
switching fabric in Figure~\ref{fig:general-model}, which are common 
components for both IP and FIA routers, and treat the power consumption 
of these components as a baseline for our analysis.
Because precise power analysis of all the components is impractical, we
make the simplifying assumption that the baseline power consumption of
FIA routers is the same as that of an IP router.

We denote the total power consumed by an IP or FIA router to forward packets as 
$P^{arch}$, the power consumption of local content caching system as 
$P_{cc}^{arch}$, the power consumption of making forwarding decisions 
as $P_{fwd}^{arch}$,  and the baseline power consumption of all the 
other components as $P_{base}$. The super-script ``$arch$'' can be substituted 
by \ip, \ndn, \scion, or \nebula. Thus, $P^{arch} = P_{base} + P_{fwd}^{arch} + P_{cc}^{arch}$.
All power consumption is measured in Watts.

\subsection{Modeling Baseline Power Consumption}
\label{sec:sub_base}

To model the baseline power consumption ($P_{base}$), we adopt the methodology 
of Lee et~al.~\cite{Lee2011} which is capable of deriving $P_{base}$ for 
heterogeneous Internet routers in core and access networks. We assume that 
FIA routers use the same technology for the components shared with today's \ip 
routers. Thus, FIA routers consume the same amount of baseline power as their 
\ip router counterparts.

We denote the power consumption of a router when idle as $P_{idle}$, the power 
consumption inscribed on the nameplate of the router as $P_{N}$, the maximal 
throughput as $I_{max}$ and the actual throughput as $I$. We can express 
$P_{base}$ as $P_{base} = P_{idle} + \alpha (P_{N} - P_{idle})$,
where $\alpha =  \frac{I}{I_{max}}$ is a factor characterizing the link utilization.
A general observation is that a core router is 
more power efficient than an edge router.
For example, a core router CRS-1 has nameplate power 16.8~kW, and 6.40~Tbps 
throughput~\cite{CiscoRouters}. In contrast, an edge router ARS-1013's nameplate 
power is 4.0~kW, but only has 0.28~Tbps bandwidth.

\subsection{Modeling Forwarding Decision Making \\Power Consumption}
\label{sec:sub_rt}
\subsubsection{Modeling $P_{fwd}^{\ip}$ for \ip routers}
Various methods exist for an \ip router to find the next interface 
to forward a packet~\cite{ruiz2001survey}. In this paper, we consider the common
hardware-based approach using Ternary Content-Addressable Memory (TCAM). 
A TCAM can perform a longest-prefix match over the entire 
routing table with a single access. However, it is known to consume at least 
three times more energy than Static Random-Access 
Memory (SRAM)~\cite{perino2011reality} and has a larger chip size. 
TCAMs are used in commodity routers such as the Cisco Catalyst 
6500~\cite{ciscocatalyst6500}.

To model the power consumption of a TCAM-based routing table, 
we assume that the power consumption of the TCAM is 
related to its size and its lookup rate.
Let $E_{TCAM}$ be the power consumed by TCAM per bit per lookup, $s$ be the size of  
a prefix record,  $N$ be the number of all prefixes stored in the FIB,
and $r$ be the average number of packets processed.
We can express the $P_{fwd}^{\ip}$ of an \ip router as 
$P_{fwd}^{\ip} = r\cdot  s \cdot N \cdot E_{TCAM} \label{eq:rt-ip-tcam}$.

\subsubsection{Modeling $P_{fwd}$ for \ndn routers}
TCAMs are inadequate to accommodate the Forwarding Information Base 
(FIB) for \ndn routers \cite{perino2011reality}. As suggested by Perino 
\etal~\cite{perino2011reality}, we analyze a scheme called ``Longest Prefix 
Match with Bloom Filters'' (LPM-BF~\cite{dharmapurikar2003longest}) as the FIB 
lookup method for a \ndn router instead of TCAM-based lookups. 

LPM-BF uses Bloom Filters stored in on-chip SRAM for the task of longest-prefix 
match in line cards. Compared with longest prefix matching using TCAM, LPM-BF 
only requires SRAM and DRAM, which are cheaper, smaller in chip size, 
and larger in capacity. Table~\ref{tab:tech} shows the power consumption of
different storage mediums. It demonstrates the advantage
of using SRAM and DRAM instead of TCAM.

\begin{table}[!b]
\centering
\begin{tabular}{cccc}
Technology & Power (Watt/bit) & Max. Size & Typical Frequency\\\hline
TCAM & $\sim3\mu W$ & $\sim$ 32Mb & $\sim$ 360MHz \\
SRAM &  $\sim40nW$  & $\sim$ 200Mb & $\sim$ 633 MHz\\
DRAM & $\sim250pW$ & $\sim$64GB & $\sim$ 1333 MHz\\
Flash &  $\sim0.3pW$& $\sim$2TB & N.A.
\\\hline
\end{tabular}
\caption{Overview of storage technologies.}
\label{tab:tech}
\end{table}

In the LPM-BF scheme, the FIB is organized by a hash 
table and stored in off-chip DRAM. Bloom filters, each of which is responsible 
to test matches for prefixes with a specific length, are stored in on-chip SRAM.
For each address, all possible prefixes are simultaneously matched against the 
Bloom filters until a longest-prefix match is found. Then the FIB hash 
table in DRAM is used to find the next hop for the matched prefix.

We divide the power consumption of an ASIC implementing LPM-BF into two 
parts: computation and storage. For computation, we primarily 
consider the computation for the Bloom filters, denoted as $P_{c}^{LPM-BF}$ in 
ASICs. For storage, we compute the power consumption of the SRAM and 
the DRAM required by the longest prefix match tasks in typical line cards, 
denoted as $P_{s}^{LPM-BF}$. We describe the power 
consumption for forwarding decisions for \ndn routers as
$P_{fwd}^{\ndn} = P_{c}^{LPM-BF} + P_{s}^{LPM-BF}$.

Let $B$ be the number of Bloom filters, $M$ be the total number of bits in the on-chip SRAM, and $N$ be the number of all prefixes. According to Dharmapurikar et~al.~\cite{dharmapurikar2003longest}, one basic configuration satisfies
that $k = \frac{M}{N} \ln{2}$ and $f = \left(\frac{1}{2}\right) ^{k}$
where $k$ is the number of hash function, and $f$ is the false-positive rate.

Let $E_{hash}$ be the power consumed to 
compute a hash function, $r$ be the number of packets requiring longest prefix 
matching per second. $P_{c}^{LPM-BF}$ can be expressed as
$P_{c}^{LPM-BF} \sim \left( Bk  + Bf + 1 \right) r \cdot E_{hash}$,
where $Bk$ is the number of hashes mandated by the Bloom filters, $Bf$ is the number of hashes caused by the false positives of the Bloom filters, and the additional hash is required to index the hash tables in DRAM. 
A typical FPGA bitcoin miner today consumes 1 Joule per 
20 MHash~\cite{bitcoinminer}. 
Thus, we choose $E_{hash}=50nJ/Hash$.

Let $s$ be the size of one record in the FIB, $\beta$ be the load factor of the 
hash table in DRAM for FIB. The size of DRAM used $O = \frac{S\cdot N}{\beta}$.
Let $E_{SRAM}$ be the power consumption of SRAM per bit per access,
 and $E_{DRAM}$ be the power consumption of DRAM per bit and $r_{max}$ is 
 its maximum frequency.
We assume that 
$\alpha$ is the proportion of the power consumption by DRAM accesses, e.g., read and write.
 $(1-\alpha)$ is the proportion of DRAM's activation and background power consumption. A typical value of $\alpha$ for DDR3 DRAM is 46\%~\cite{drampower}.
$P_{s}^{LPM-BF}$ can be 
expressed as
$P_{s}^{LPM-BF} = r \cdot M \cdot B\cdot k \cdot E_{SRAM} + \frac{r \cdot (Bf + 1)}{r_{max}} \alpha \cdot O \cdot E_{DRAM} + (1-\alpha) O \cdot E_{DRAM} $.
$r \cdot M \cdot B\cdot k \cdot E_{SRAM}$ stands for the power consumed by the on-chip SRAM.
$\frac{r \cdot (Bf + 1)}{r_{max}} \alpha \cdot O \cdot E_{DRAM}$ stands for the power consumed by DRAM accesses,
and $(1-\alpha) O \cdot E_{DRAM}$ is DRAM's activation and background power consumption.

\subsubsection{Modeling $P_{fwd}$ for \nebula and \scion routers}
Unlike \ip and \ndn, both \nebula and \scion use packet-carried state for 
finding the interface to forward a packet. In other words, the forwarding 
decisions resides in the packet header and no route table needs to be stored on 
routers. The lack of routing tables (and thus lack of relatively expensive 
table lookup operations) helps reduce the power consumption of packet 
forwarding. However, both \nebula and \scion routers use cryptographic 
primitives to verify the integrity of the routing decisions embedded in the 
packet headers, which add to $P_{fwd}$. 

Since the verification of the routing decisions carried in packets is the only 
computation-intensive operation in the forwarding process for \nebula and 
\scion, we only consider the computation of cryptographic verification when 
modeling $P_{fwd}$. Let $E_{verif}^{FIA}$ be the power consumed to verify the 
routing decision carried in the packet. We express $P_{fwd}$ as 
$P_{fwd}^{FIA} = r \cdot E_{verif}^{FIA}$.

In \nebula, the verification process involves verifying the 
``Proof of Consent'' (PoC) and ``Proof of Provenance''  (PoP) carried in 
the 
packets as well as generating new PoPs to prove provenance. Let $l_{AS}$ be 
the average AS-level path length. The average energy consumed on a \nebula 
border router to verify packet-carried routing decisions $E_{verif}^{\nebula}$ 
can be expressed as 
$E_{verif} ^ {\nebula} = E_{hash} + (l_{AS}^2 + l_{AS} + 2) E_{AES}$.
We choose $l_{AS}=4.4$  as measured by 
Kuhne and Asturiano~\cite{as_path_len}.

In comparison, the verification process on a \scion border router requires only 
one AES-MAC computation to verify that the HF was generated by 
the border router itself. As a result, the energy consumed by a \scion border 
router to verify the packet-carried state $E_{verify}^{\scion}$ can be 
expressed as
$E_{verif}^{\scion} = E_{AES}$.
A single 128-bit AES operation on an Intel CPU with AESNI technology 
consumes 4.8 cycles/byte on i7-980X with frequency 3.3GHz and 
12 threads~\cite{AESNI}. The maximal power consumption of i7-980X CPU is 130W. 
Accordingly, we choose $E_{AES} = 250nJ/AES op$.

\begin{figure}[htbp]
  \centering
  \includegraphics[width = 9cm]{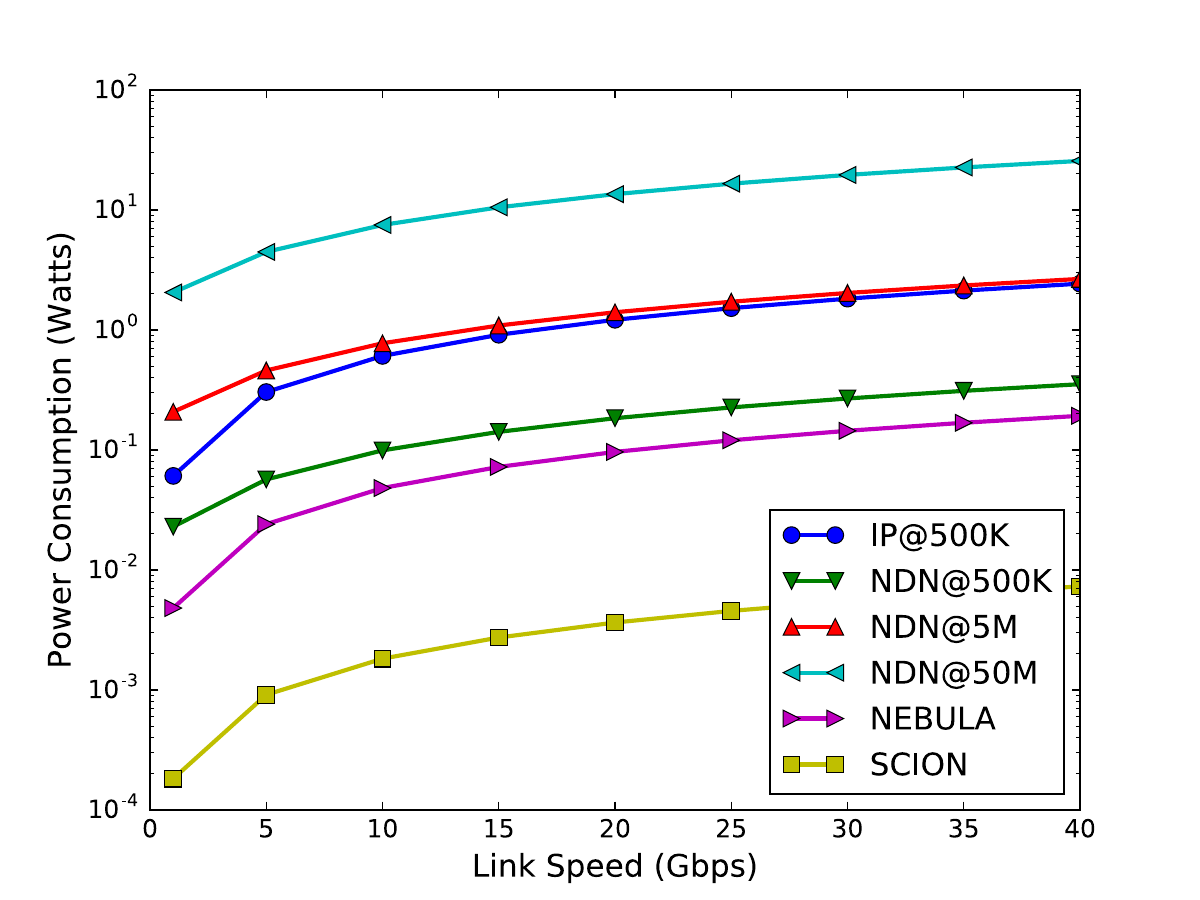}
  \caption{$P_{fwd}$ under different link speed for border routers. For \ndn, we evaluate routing tables containing 500K, 5M, and 50M entries.}
  \label{fig:sig-ls-archs}
\end{figure}

\textbf{Comparison of $P_{fwd}$ among FIAs}
Figure~\ref{fig:sig-ls-archs} 
graphs the power consumption of forwarding decision making for a line 
card plugged in a border router of different FIAs when link speed varies from 
1Gbps to 40Gbps. 
For the FIB size of \ndn, Perino and Varvello have suggested 20 million 
entries~\cite{perino2011reality}. Accordingly, we vary \ndn{}'s FIB size 
from 500K entries to 50M entries to demonstrate the influence of routing-table
size on routers' power consumption. Because we assume that the power consumption of a router's TCAM
 is only related to the TCAM's size, the power consumed by an \ip router is 
constant when link speed changes.

In general, making forwarding decisions using packet-carried state consumes less
power than that using routing table lookup. Depending on the complexity of the 
packet-carried state verification, the power consumed can vary 
significantly. In \nebula which requires several crypto-operations to 
forward each packet, the performance improvement over \ip is 1 order of magnitude.
 Since \scion
only requires a single AES-MAC to verify the packet-carried state, the 
performance of making a forwarding decision is more than 3 orders of magnitude more 
efficient than that of an \ip border router. 

On the other hand, because LPM-BF consumes less power at low link speed,
routing table lookups in \ndn with 500K records in the route table consumes 10X 
less power than that using TCAM in an \ip border router. However, the 
advantage becomes less apparent as the routing table size grows. As a matter of fact, the routing-table lookup using LPM-BF in \ndn 
at 40Gbps with 50M records in each route table consumes up to 12 times 
more power than that using TCAM in an \ip border router.

\subsection{Modeling Content Caching Power Consumption}
\label{sec:sub_cc}
\subsubsection{Edge caching $P_{cc}$}
To estimate the power consumption of edge caching, we consider a
state-of-the-art of content cache from Netflix~\cite{netflixopenconnect}. 
Netflix is the 
largest video distribution service in the world, accounting for as much as 
34\% of worldwide network traffic during peak hours~\cite{Sandvine2013}. As one 
of the leaders in the space, Netflix has strong incentives to use powerful 
yet power-efficient hardware. Netflix's content routers have storage 
capacities of 108TB, support 10Gbps network throughput and consume 
600 Watts~\cite{netflixopenconnect}. For comparison purposes in this paper, we 
use 600W/10Gbps as our baseline $P_{cc}$ for networks with edge caching. 

\subsubsection{Pervasive caching $P_{cc}$}
In current FIAs, the design and implementation of content caches 
(particularly caches in high-speed core routers) remains an open problem. 
Content cache systems in FIAs generally have the following design goals: 
1) rely upon inexpensive hardware to encourage massive deployment; 
2) allow provisioning of higher storage capacity as link speeds increase; 
and 3) serve contents at high speeds, ideally close to the arrival rates of 
packets. 

Since pervasive caching routers are still a relatively new concept, measurement 
data on power consumption of real-world devices is, to the best of our 
knowledge, not yet available. Thus, we consider two key-value object store schemes 
(HashCache~\cite{badam2009hashcache} and SILT~\cite{lim2011silt}) that fulfill 
the design requirements. We envision content cache in FIAs to use HashCache 
or SILT, whichever consumes the least power. 

\textbf{Key-value store.} Both HashCache and SILT implement a two-layer 
architecture: 1) an underlying storage layer in large and relatively 
slow medium storing the actual objects, and 2) an indexing layer in a small and 
relatively fast medium to efficiently handle the queries and locate the corresponding
content. A typical setup for these key-value stores includes an SSD-based 
storage layer indexed by a DRAM-based indexing layer.

There are three key parameters characterizing the key-value store schemes: 
$\kappa$, determining the number of bytes needed in the index layer for each 
object in the storage layer, and two amplification factors, $A_{rd}$ and 
$A_{wr}$, determining the number of reads or writes required for the storage 
layer when there is a read hit or a write hit in the storage layer. 
Table~\ref{tab:key-param-cs} lists the typical values for both two key-value 
store schemes.

\begin{table}[!b]
  \centering
  \begin{tabular}{cccc}
    \hline
    Method & $\kappa$ (Bytes/Object) & $A_{read}$ & $A_{write}$\\\hline
    SILT & 1 & 1.01 & 4 \\
    Hashcache(SetMem) & 11/8 & 1 & 1 \\
    Hashcache(logLRU) & (15$\sim$47)/8 & 1 & 1 \\\hline
  \end{tabular}
  \caption{Key parameters for content store algorithms.}
  \label{tab:key-param-cs}
\end{table}

We use a conservative estimate about the power consumption of key-value stores 
by only accounting for the power consumption of the underlying storage medium. 
We leave out the power consumed by hash computations needed, because the 
hash functions can be computed very efficiently~\cite{hashfunccycles}. 
\chen{what we should do for the power consumed by rebuilding the key-value 
store in SILT? What about other computations?}

Let $C_{st}$ be the storage capacity of the storage layer, $E_{st}$ be the 
power consumed to store each bit by the storage medium supporting the storage 
layer, and $E_{idx}$ be the power consumed to store each bit by the storage 
medium implementing the index layer. We  derive
$P_{cc} =  \left(E_{st} + \kappa E_{idx} \right)C_{st}$.

\textbf{Choosing storage mediums.}
Selecting storage mediums for both the storage and index layers involves 
considering both the storage capacity and transaction rates desired for the 
key-value store and those offered by current technologies. Specifically, we 
take into account two categories of limitations: 
\emph{storage-capacity limitation} and \emph{transaction-rate limitation}. 

Let $(C_1^{max}$, $R_1^{max})$ be the maximal storage capacity and maximal 
transaction rate for the storage medium used by the index layer, and 
$(C_{st}$, $R_{st})$ be those for 
the storage medium 
used by the storage layer. We express the \emph{storage-capacity limitation} 
by: 1) $\kappa C_{st}  \le  C_1^{max} $, and 2) $C_{st}  \le C_2^{max} $.

Let $\lambda_{in}$ be the arrival rates of content-distribution-relevant packets, $\alpha$ be the percentage of the content queries (the other packets are data packets), $r_{hit}$ be the percentage of cache hit rate for the content queries, $r_{mod}$ be the probability of writing to add new cached objects. We express the \emph{transaction-rate limitation} as: 1) $\lambda_{in}  \le  R_1^{max}$, and 2) $\left(\alpha r_{hit} A_{read}+ \left(1-\alpha \right) r_{mod} A_{write} \right) \le R_2^{max}$.
\begin{figure}[tb]
	\centering
	\includegraphics[width=.45\textwidth]{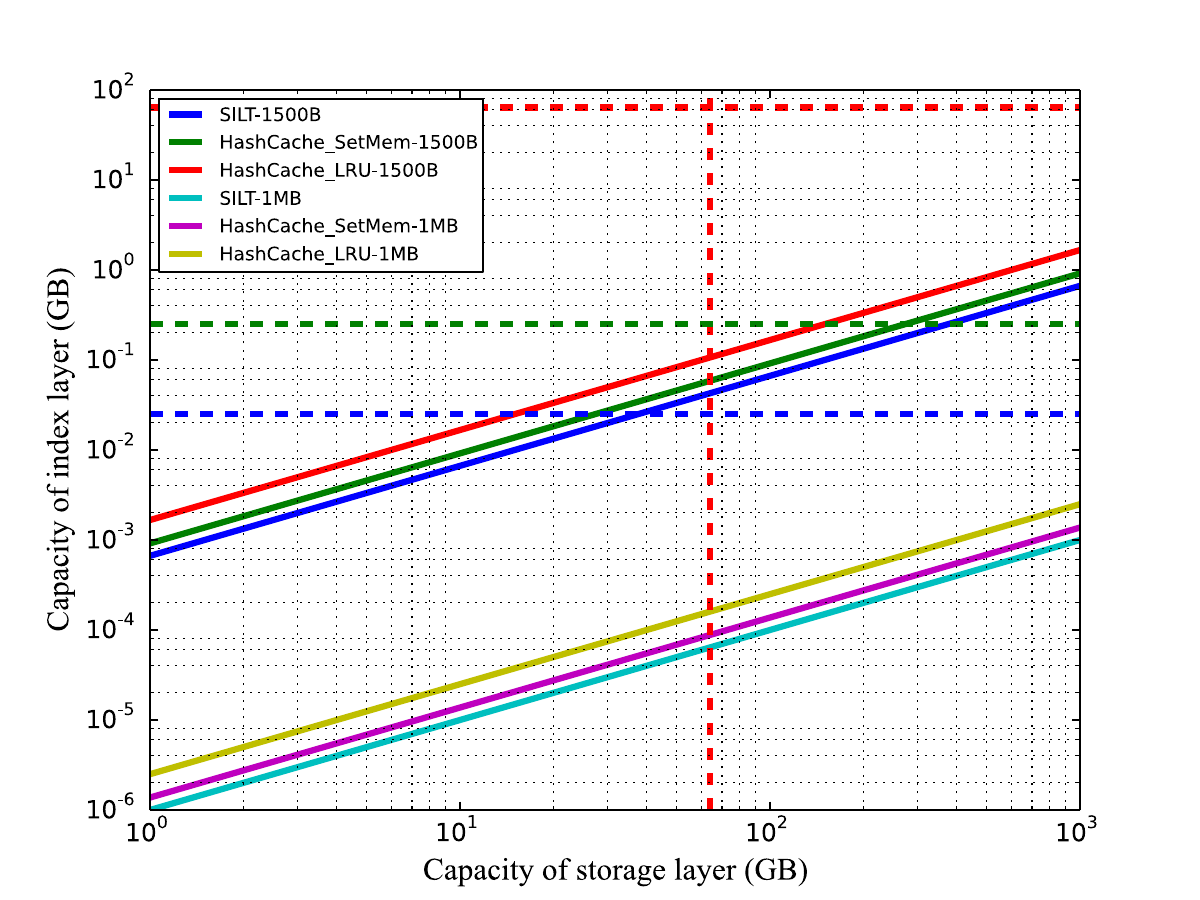}
	\label{fig:index_vs_storage}
	\caption{The size of the index layer as a function of the size of the storage layer for different key-value stores. The blue, green, and red dashed line shows the maximal sizes of SRAM, RLDRAM and DRAM, respectively.}
\end{figure}
\begin{figure*}[htbp]
\centering
\subfigure{
	\includegraphics[width=.45\textwidth]{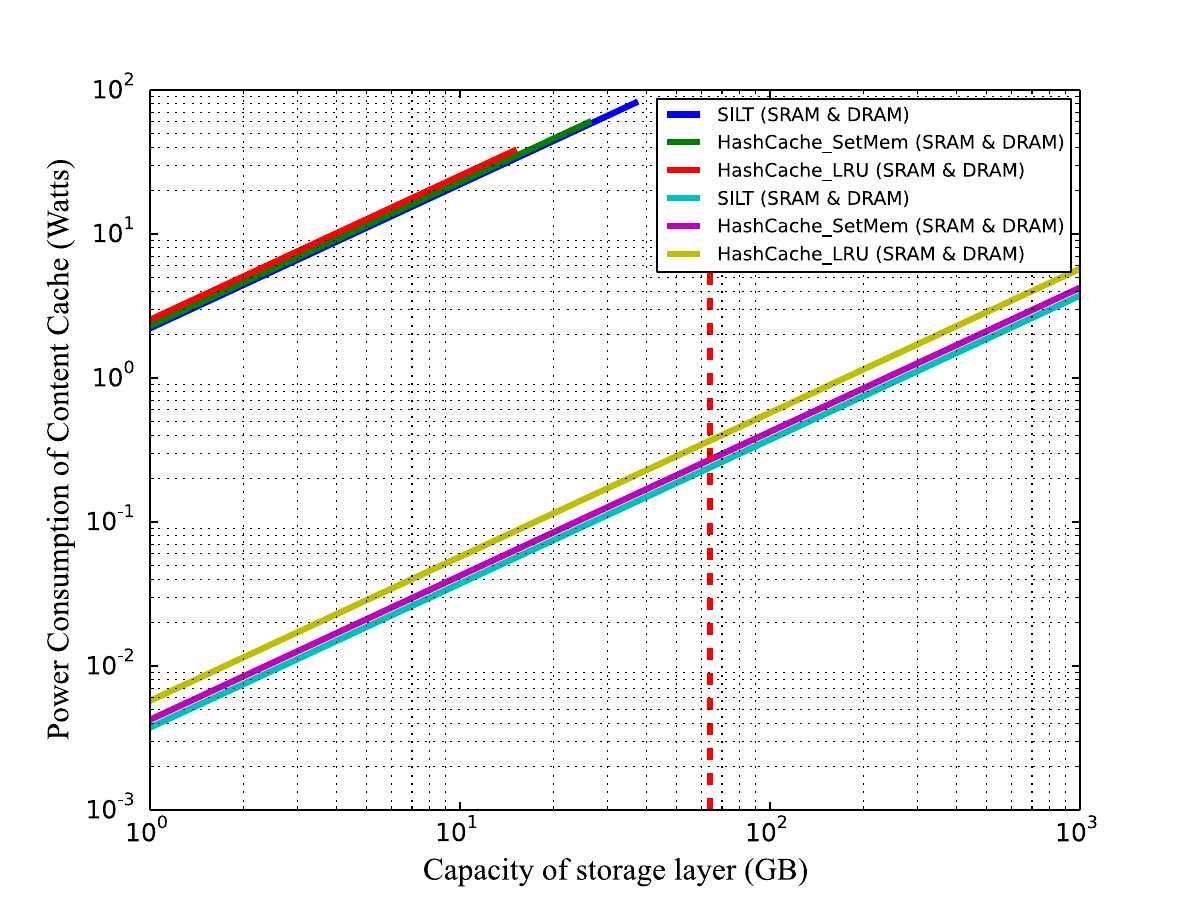}
	\label{fig:power_vs_size}
}
\subfigure{
	 \includegraphics[width=.45\textwidth]{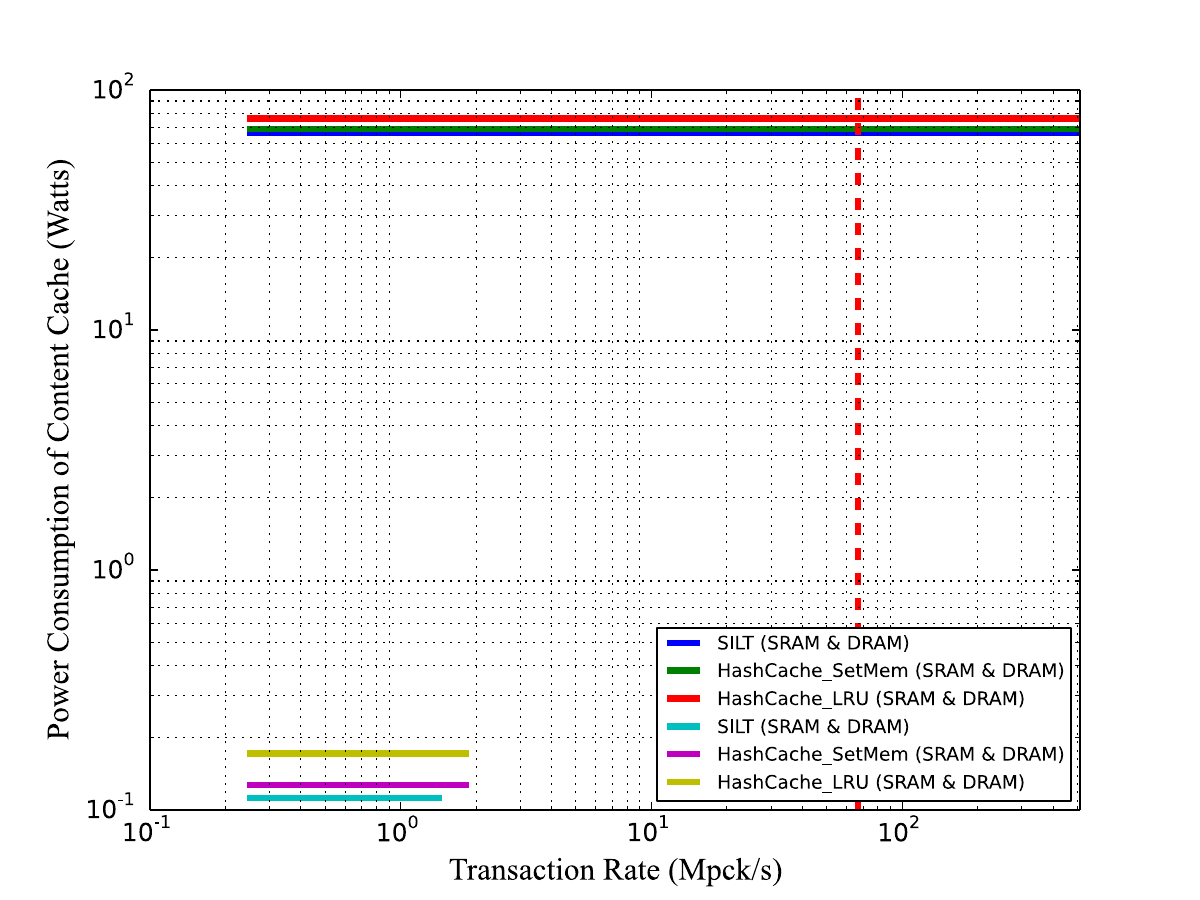}
	 \label{fig:power_vs_x_rate}
}
\caption{(a) The power consumed by content caches of various sizes. Either SRAM-DRAM or DRAM-SSD is chosen as the combination of storage mediums. The red dashed line shows the maximal size of DRAM. (c) The power consumed by content caches for different transaction rates with parameters $\alpha$=0.5, $r_{hit}$ = 0.1, $r_{mod}$ = 0.01. The red dashed line shows the maximal transaction rate of DRAM.}
\end{figure*}

Figure~\ref{fig:index_vs_storage} shows the size of the index layer as a 
function of the size of the storage layer for different key-value stores. When 
the object size is small (e.g., 1500 bytes), the size of 
the index layer is the major bottleneck to build a content cache with large 
capacity and high transaction rates. In contrast, when the object size is 
larger (e.g., 1MB) such as multimedia content, the 
size of the DRAM becomes the bottleneck.

Figures~\ref{fig:power_vs_size} and \ref{fig:power_vs_x_rate} show 
 the power consumed by content cache with different sizes and capable of 
handling different transaction rates. In general, an SRAM-DRAM combination 
could offer 300X higher transaction rates while consuming 600X more power than a
DRAM-SSD combination. For a router with high link speed ($\ge$ 10Gbps), 
an SRAM-DRAM combination could be leveraged to implement the content cache, 
while DRAM-SSD combination is suitable to implement a power-efficient content 
cache for a router with low link speed ($\le$ 2Gbps).

As for the comparison between different key-value stores, SILT benefits from its 
smaller value of $\kappa$ in two aspects. First, SILT has a smaller index layer. 
As the index layer is more power 
hungry, this feature renders SILT to be 3\% and 15\% more efficient compared to 
HC-SetMem and HC-LogLRU, respectively. Second, when implementing 
packet-level caches using the SRAM-DRAM combination, SILT could implement a 
content cache with a 1.3 to 2 times larger storage layer. 

On the other hand, with larger amplification factors, the transaction rate handled by SILT is lower when the storage layer speed becomes the bottleneck. This is the case when DRAM-SSD is chosen to implement content cache of larger capacity, where the maximal transaction rate of SILT is 25\% lower than those of HC-SetMem and HC-LogLRU.

\begin{figure}[!b]
	\centering
	\includegraphics[width = 9cm]{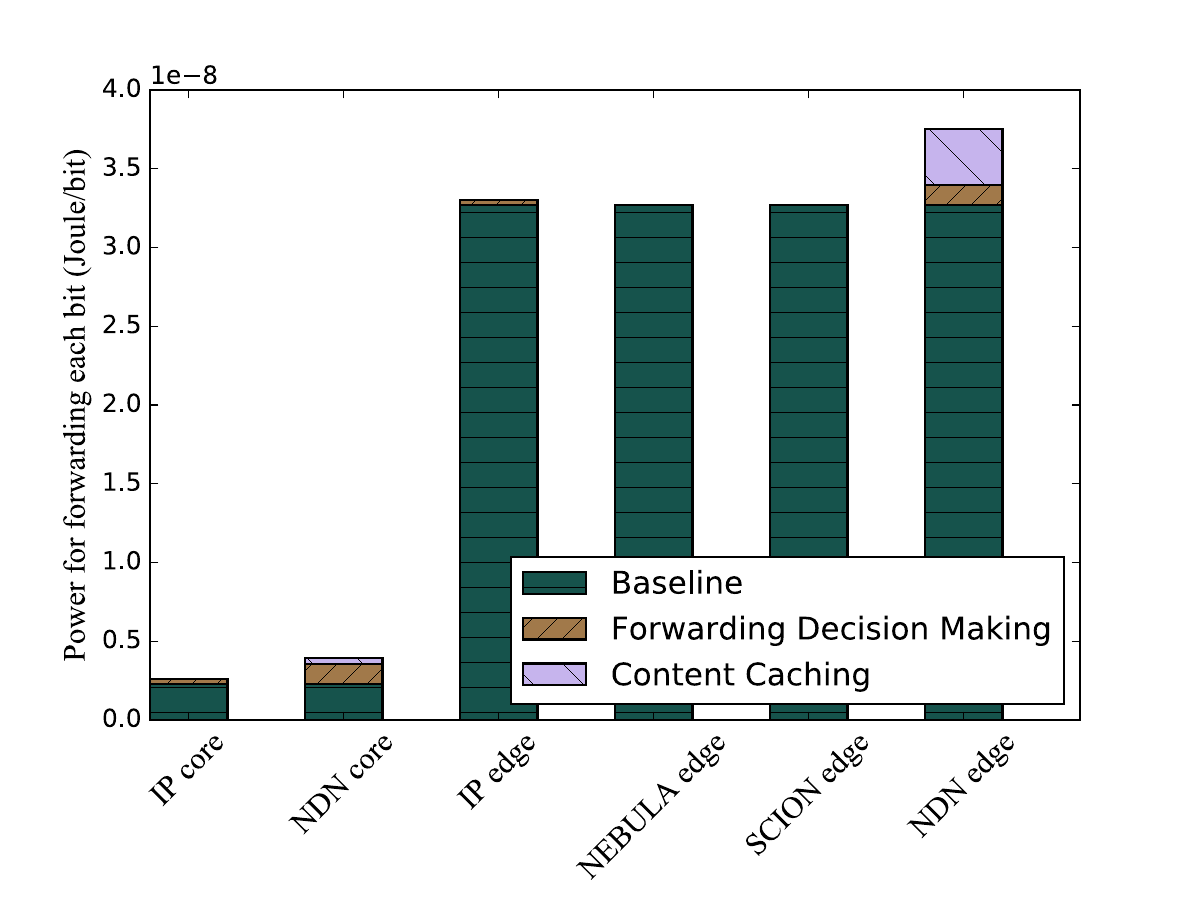}
	\caption{Power consumption of FIA core and edge routers. 
		``core'' means core routers, and ``edge'' means edge routers. We omit \scion and \nebula core routers' results as forwarding decision making only happens on edge routers. }
	\label{fig:power_fia_router}
\end{figure}

\subsection{Summary} 
With the models constructed in Sections~\ref{sec:sub_base}, \ref{sec:sub_rt}, 
and \ref{sec:sub_cc}, we can compute the power consumed to forward packets for 
each type of FIA routers. To normalize the results, we calculate the energy 
used for forwarding a single bit on each FIA 
router. 

Figure~\ref{fig:power_fia_router} shows the Joule per bit for each 
FIA router with a specific configuration. For \ndn, each core router is 
equipped with 1TB DRAM-based content cache managed by SILT with SRAM as the 
index layer and each edge router is equipped with 256GB content cache using
the same key-value store setup.
 \ip, \nebula, and \scion routers do not perform any content cache.
All interfaces of \ndn routers use ASICs based on LPM-BF to make forwarding 
decisions. We assume the number of entries in \ndn{}'s FIB is 20M and
the number of entries in the \ip{} routing table is 512K.

According to Figure~\ref{fig:power_fia_router}, \ndn routers consume 
more power than \ip and other FIA routers due to their 
content cache and larger FIB. The increase in the power budget for a \ndn router 
ranges from 72\% for core routers to 15\% for edge routers compared to a \scion 
router, which consumes the least power due to its efficient forwarding.

Though \scion and \nebula are more power efficient than \ndn when 
we only consider traffic forwarding, content caches may reduce the overall 
number of bits that are transmitted. In the following section, we further 
explore this trade-off by performing large-scale simulations.
 
\section{Simulation}
\label{sec:simulation}\label{sec:sim}
Based on the model of FIA routers in Section~\ref{sec:model}, we now
 compare the power consumption of each 
FIA in content distribution scenarios. We conduct our experiments by 
simulating the forwarding behaviors of the \ip network and FIAs when 
used for content distribution.

\subsection{Simulation Setup}

\begin{figure}[t]
\centering
\subfigure[]{
	\includegraphics[scale=0.8]{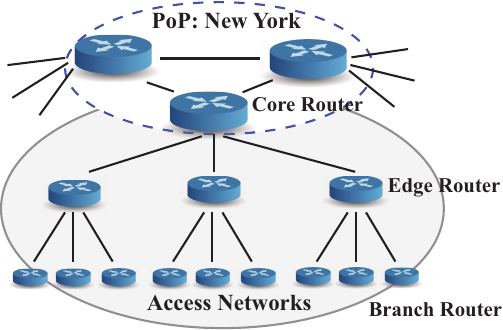}
	\label{fig:topology_pop}
}
\caption{Access tree connected to the New York PoP. The access tree shown is a 
complete tree with depth=3, arity=3.}
\end{figure}
 
\textbf{Topology.} The topology used in our simulations is based on 
education backbone networks and the Rocketfuel topology~\cite{spring2002measuring}. 
We obtain router-level information for two education/research backbones: Geant
and Abilene. We also extract router-level information for six different ISPs: 
Telstra (AS1221), Sprint (AS1239), NTT (AS2914), Verio (AS3257), 
Level3 (AS3356), AT\&T (AS7018).
Routers are grouped according to their Points of Presence (PoPs). The PoPs are 
then annotated with city-level location information.

We follow the methods proposed by Fayazbakhsh et al.~\cite{fayazbakhsh2013less} 
to approximate access networks by 
trees appended to each PoP. The internal nodes of the trees are edge routers. 
We use complete trees with varying depths and arities. 
\chen{analyze influence of different parameters}

\textbf{Traffic Patterns.}
Content distribution traffic
represents a significant amount of overall Internet traffic. Thus, our primary 
goal is to evaluate the power consumption of FIAs in content 
distribution scenarios. 
For traffic access patterns, previous work has suggested 
 that a Zipf distribution closely approximates real world content 
access from end hosts~\cite{fayazbakhsh2013less}. The key parameter
$\alpha$ in a Zipf distribution decides the relative popularity of different 
contents. A larger $\alpha$ means that popular content queries constitute
a larger proportion of all queries, which also means more 
temporal locality in the content access pattern.

 We use synthesized 
content-access traces with $\alpha=0.99$ (which approximates US users' 
behaviors) as suggested by Fayazbakhsh et al.~\cite{fayazbakhsh2013less}. 
For the query distribution, we employ the population of the city for each PoP to 
distribute the queries across the access networks belonging to each PoP. For simplicity, the queries are assumed to only enter the network through leaf nodes of each access network.

\subsection{Routing-table lookup versus packet-carried state}
We begin by evaluating the power consumption with respect to the first design 
choice (\ie routing-table lookup versus packet-carried state) for making 
forwarding decisions. For architectures leveraging routing-table 
lookups, we consider current \ip networks and \ndn. For those using
packet-carried state, we consider SCION and \nebula. To prevent measurement 
noise induced by caching, we intentionally remove the content caching module 
for \ndn in the simulation. We will add it back in Section~\ref{sec:sub:cache}. 

As discussed in Section~\ref{sec:sub_rt}, the primary sources of routing 
table power consumption are routing table maintenance and 
routing table lookup operations. We assume TCAMs are used for the \ip network 
and LPM-BF is used for \ndn, because LPM-BF can search a larger \ndn routing
table and consume less power.
In the case of \ip networks, we set the number of prefixes to be 
500K according to the FIB size for BGP from RouteViews~\cite{routeview}. In case 
of \ndn, we choose the number of (content name) prefixes to be 20M which can 
be supported by a 200Mbit on-chip SRAM. 

The main sources of power consumption for packet-carried state forwarding 
are state verification and transmission of extra bits in packet 
headers. For the verification of packet-carried state, we assume 128-bit AES 
as the pseudo-random permutation to construct multiple crypto-primitives 
in \nebula and \scion. For the packet payload size, we select 1350 
bytes~\cite{saroiu2002analysis} for content responses and 40 bytes for content 
queries.

We focus on inter-domain forwarding 
decisions because BGP routing tables are several orders of magnitude larger 
than intra-domain routing tables. Furthermore, we assume 
only the network layer protocols differ among FIAs, but all the other layers 
(transport layer, application layer, etc.) in the network stack remain the same.

\begin{figure}
\includegraphics[width=9cm]{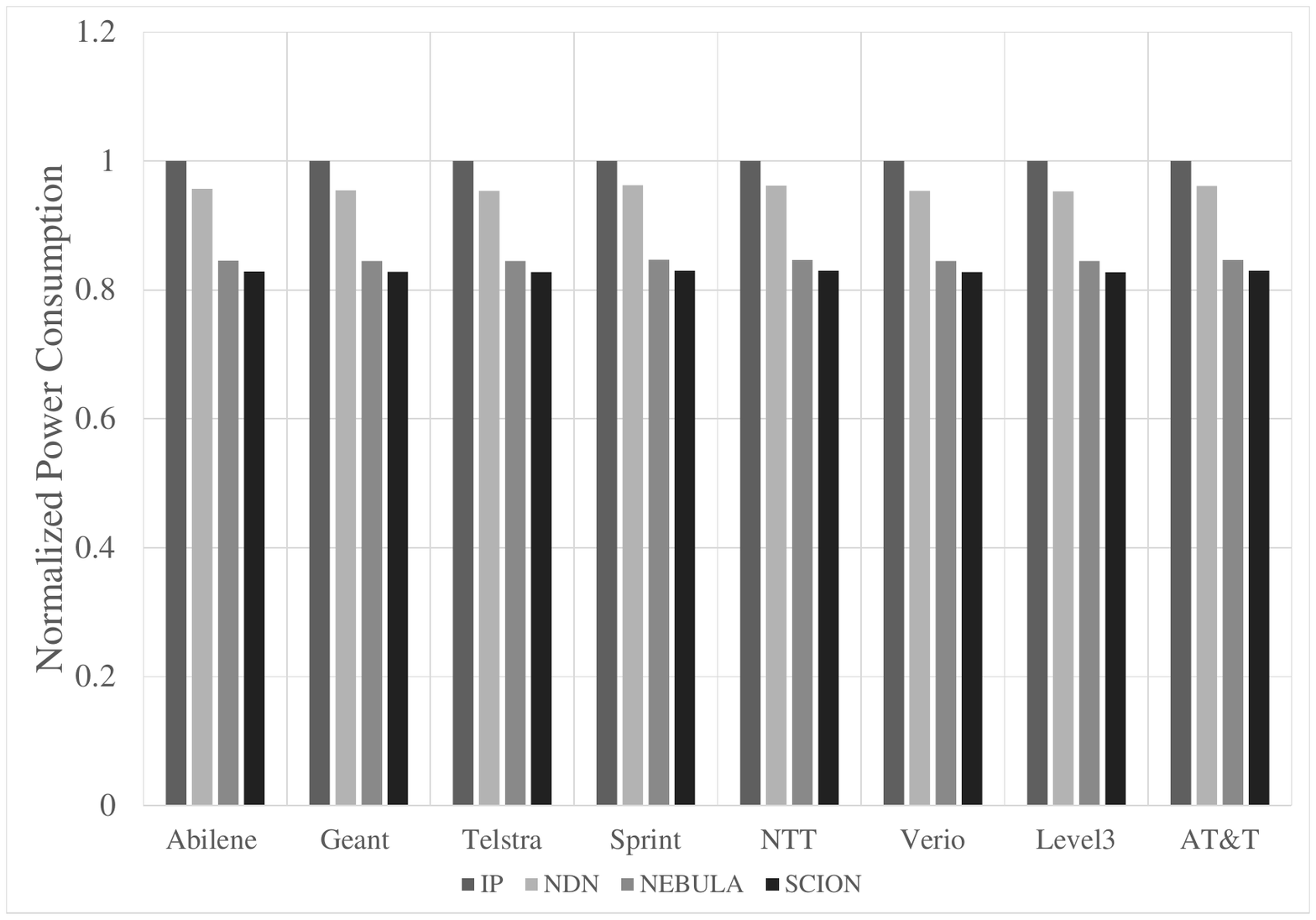}
\centering
\caption{Power consumption of \ip, \ndn, \nebula and \scion. There is no edge 
cache in the network. For each topology, the results are 
normalized by baseline results of the \ip network.}
\label{fig:sim_result_0} 
\end{figure}

Figure~\ref{fig:sim_result_0} demonstrates the power 
consumed by the synthesized content-access traces in different AS topologies. 
Across all AS 
topologies, \emph{making forwarding decisions by using packet-carried state 
is 15\% more efficient than by doing routing table lookups}. The reason, as 
partially described in Section~\ref{sec:sub_rt}, is that maintaining and 
searching routing-tables on individual routers consumes more power than 
including forwarding decisions within each packet. 

Next, we compare FIAs using the same forwarding method. 
For \ip and \ndn, both of which use routing-table lookup, \ndn routers
consume 4\% less power than \ip due to efficent forwarding-decision making. 
Though we assume \ndn routers have larger routing tables that contain 
20M entries each, \ndn routers consume less power because LPM-BF allows
\ndn routers to store and search routing tables more efficiently.

For \nebula and \scion, both of which use packet-carried state, \scion routers
consume 3\% less power than \nebula routers.
This is partially due to packet-carried state verification being more 
expensive in \nebula routers than in \scion routers and partially due 
to \nebula's larger packet headers.

\subsection{Edge caching versus pervasive caching}
\label{sec:sub:cache}
In this section, we consider the influence of different caching methods on the 
power consumption of packet forwarding in different network architectures. 
Particularly, we consider \ip, \nebula, \scion with edge caching and \ndn which 
inherently supports pervasive caching. For completeness, we also evaluate 
\ip, \nebula and \scion without content caching. Because \scion without caching 
was shown to be the most power efficient in the previous section, we use it as 
the baseline result to normalize the results of other architectures.
While definitions of edge caching may vary, our simulation follows the one used by Fayazbakhsh et al.~\cite{fayazbakhsh2013less}, i.e., only leaf nodes in access networks cache contents.

\textbf{Cache Budget Ratio.}
Cache capacity is the primary factor that impacts cache performance and the 
power consumption of the caching device. We define cache budget ratio in order 
to fairly compare edge caching and pervasive caching. Let $R$ be the number of 
routers capable of caching, $C$ be the average cache capacity of each router, 
$O$ be the total number of individual contents in the network, and $s$ be the 
average size of each content. We define the cache budget ratio $c$ as $c = \frac{R \times C}{O \times s}$.

We choose $c=5\%$ as a baseline which is observed as a relationship between 
CDN cache provisioning and the total requested objects seen by the cache each 
day~\cite{fayazbakhsh2013less}. We assume that cache capacity is uniformly 
distributed among  router.

\textbf{Cache Replacement Strategy.} 
Cache replacement strategies are important for content routers to exploit the 
locality of content accesses. Therefore, cache replacement strategies are 
expected to influence the performance of content routers in power consumption. 
In our simulation, we select the Least-Recently Used (LRU) method as our 
baseline strategy. Note that designing an optimal or high-performance cache 
replacement strategy is out of scope for this paper. 

\textbf{Cache Discovery Strategy.}
For edge caching used in \ip, \nebula and \scion, we simply assume that content 
request is only served by each standalone cache server. Infrastructure to 
coordinate the cache servers~\cite{Akamai} is not provisioned. We call this 
strategy \emph{simple edge caching}. For pervasive caching, we assume 
\emph{on-path cache discovery}, in which only content cached in the on-path 
routers would be served.

\begin{figure}
\centering
\includegraphics[width=8.5cm]{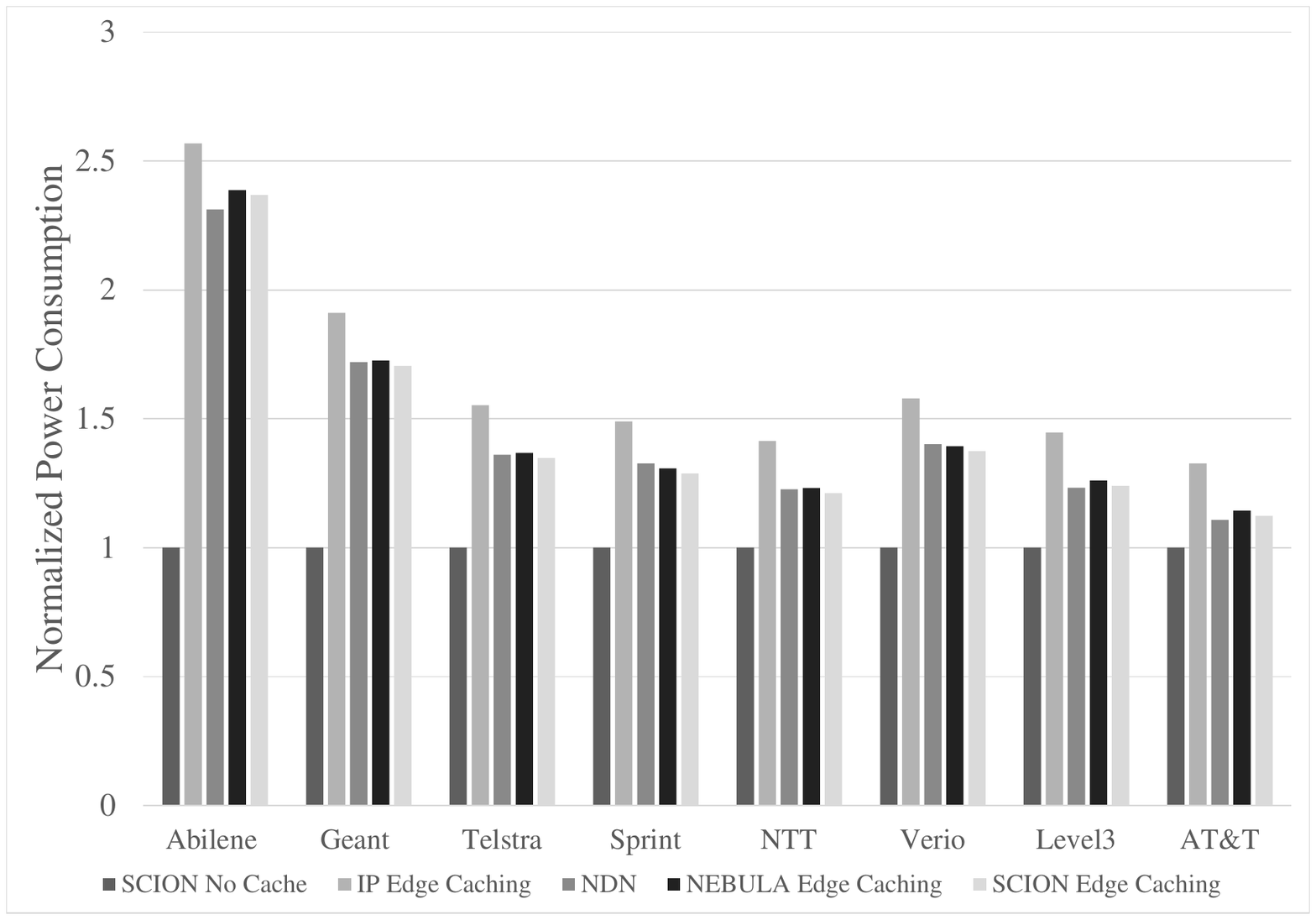}

\caption{Power consumption of end-to-end communication, edge caching and 
pervasive caching with capacity ratio $c=0.05$. All results are normalized by
\scion with no content caching.}
\label{fig:result_mid_cache}
\end{figure}

Figure~\ref{fig:result_mid_cache} shows the power consumed by different network 
architectures with caching. We use the baseline capacity ratio $c=0.05$, and we 
also evaluate \emph{simple edge caching} as the cache discovery strategy for 
networks with edge caching and \emph{on-path cache discovery} as the
 strategy for networks with pervasive caching. We use a synthesized 
content-access trace following a Zipf distribution with $\alpha=0.99$, which is 
the $\alpha$ computed from a real-world content-access 
trace~\cite{fayazbakhsh2013less}.

Surprisingly, network architectures with caching 
enabled tend to consume 15-100\% more power than those that do not cache. 
Because a smaller $\alpha$ value implies less locality in the content access 
pattern, caching becomes less efficient, which results in
caching consuming more power in the access pattern scenario of our simulation.
In other words, \emph{networks with end-to-end communication without caching
seems to consume less overall power compared to those that use caching}.

Regarding the comparison among network architectures with caching, 
\ndn with pervasive caching only saves on-average 2\% power compared to 
\scion with edge caching, which is the most power-efficient among \ip,
\nebula, and \scion with edge caching. 
Compared to \ip, which also leverages routing-table
lookup for making forwarding decisions, \ndn consumes up to 16\% less power.
The result implies that \emph{pervasive caching helps reduce power 
consumption, but the power budget cut is limited}.
The reason is two-fold: 1) multiple-layer caching or cooperative caching
provide limited improvement to single-layer caching, as indicated by
previous works~\cite{wolman1999scale, fayazbakhsh2013less}; 2) pervasive
caching requires more power-consuming caching devices, which
further reduces the small advantage in power consumption 
by having shorter length.

Finally, \emph{\scion with no caching consumes the least amount of power}. 
According to our previous analysis, \scion with no caching benefits 
from two design choices: 1) efficient verification of packet-carried state
for making forwarding decisions, 2) end-to-end design without caching. 
 
\textbf{Sensitivity analysis}.
We also conduct sensitivity analysis with respect to different cache
budget ratios $c$, different content access patterns, and multiple cache
discovery strategies. Results show that the observations remain true for
various combinations of parameters. We document our evaluation method
and results of sensitivity analysis in detail in our technical report~\cite{techreport}.
\subsection{Summary of Key Observations}
\begin{enumerate}
\item Network architectures that use packet-carried state instead of 
routing-table lookups exhibit lower power consumption. This observation holds 
even in the presence of larger packet headers. 
\item FIAs without caching consume less overall power compared 
to those that use caching. 
\item The use of pervasive caching results 
in marginal reductions in power consumption. 
\item Among the studied FIAs, \scion with no caching
consumes the least amount of power. 
\end{enumerate}

\subsection{Sensitivity Analysis}
\label{sec:sensitivity-analysis}

In order to further understand the influence of various parameters on our 
results, we conduct a one-dimensional sensitivity analysis: for each analysis,
we vary only one of the parameters while fixing all the other parameters.

\begin{figure*}[ht]
	\centering
	\subfigure[]{
		\includegraphics[width=0.3\textwidth]{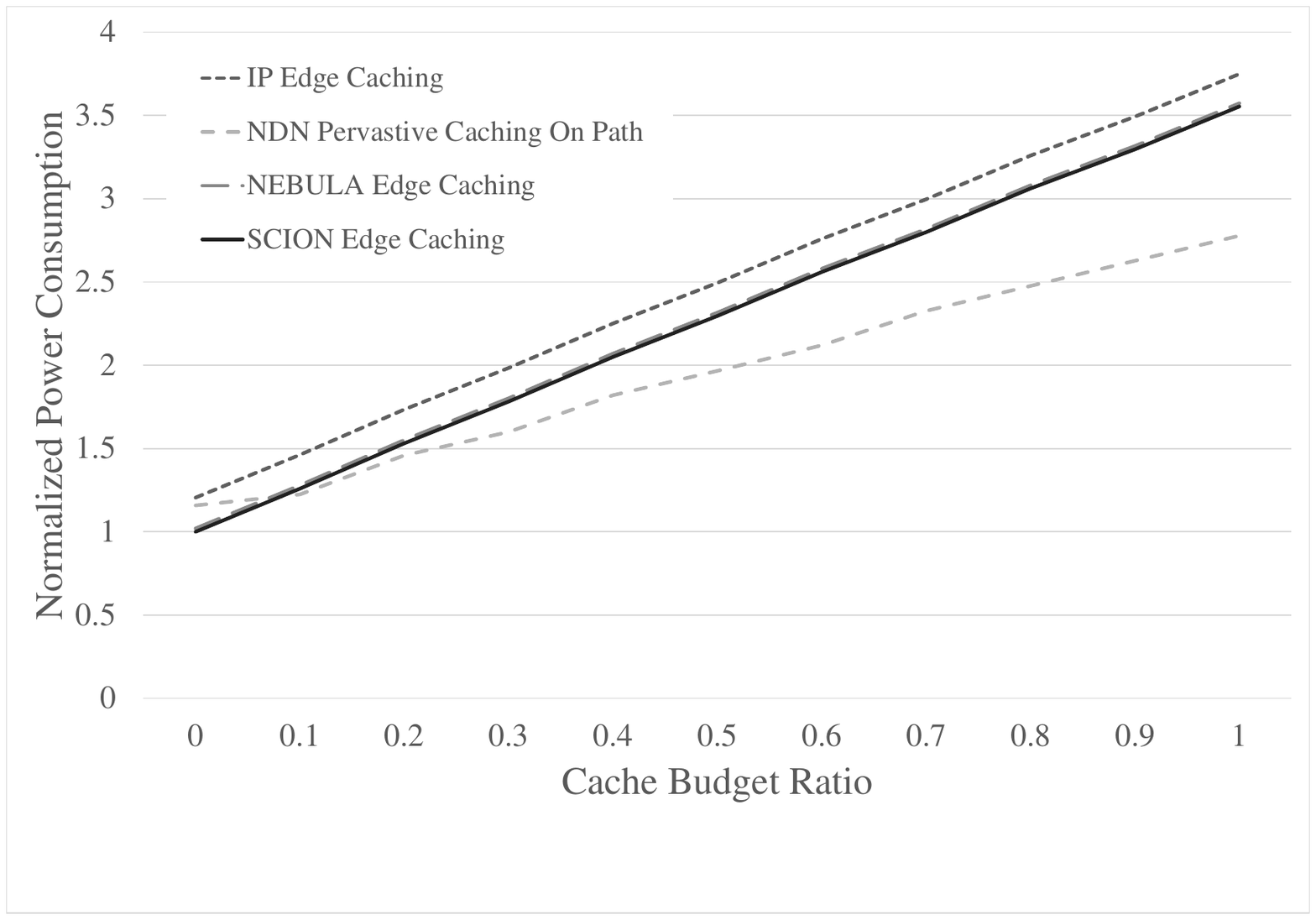}
		\label{fig:result_cache_budget}
		\centering
	}
	\subfigure[]{
		\includegraphics[width=0.3\textwidth]{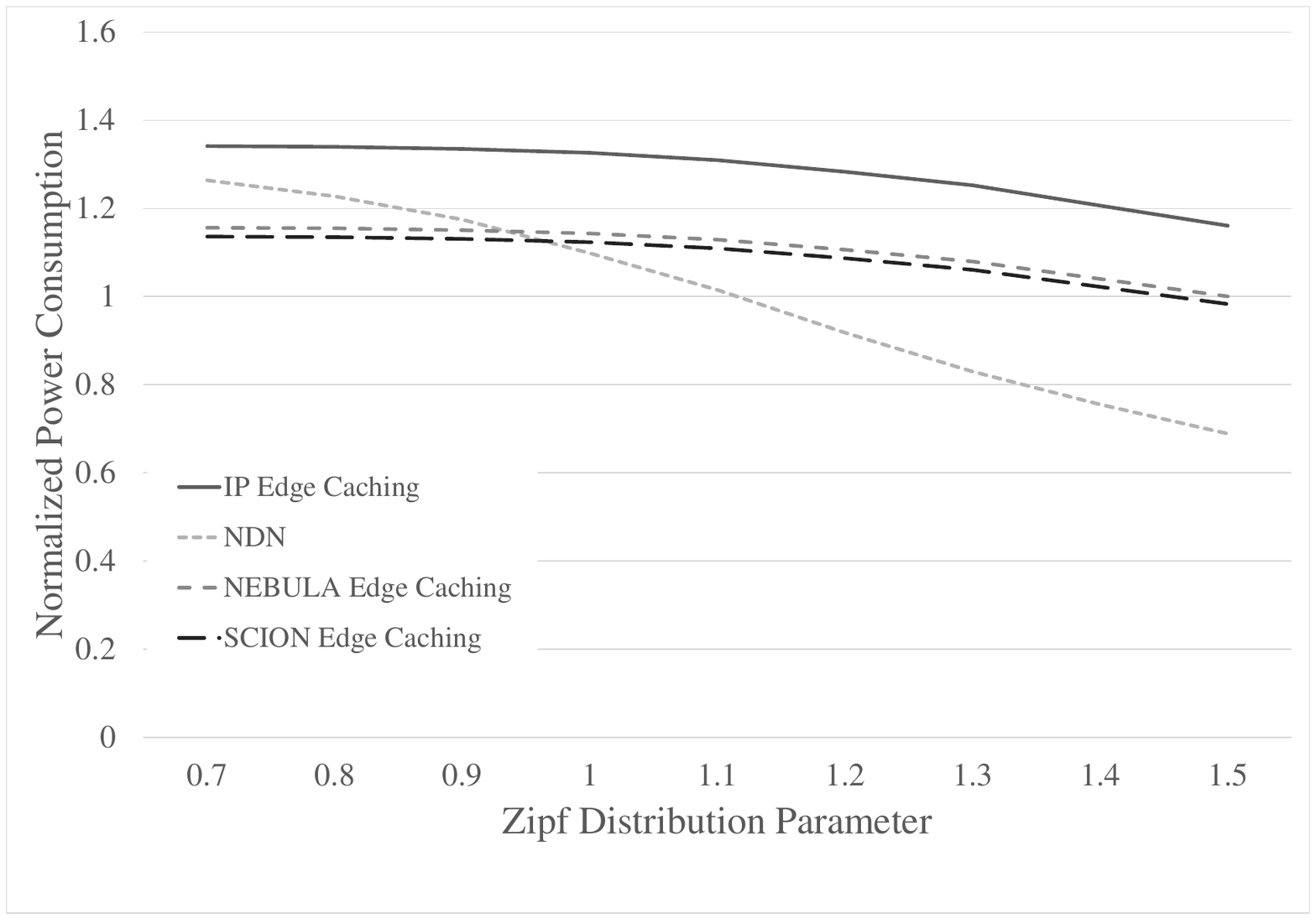}
		\label{fig:result_zipf_dist}
		\centering
	}
	\subfigure[] {
		\includegraphics[width=0.3\textwidth]{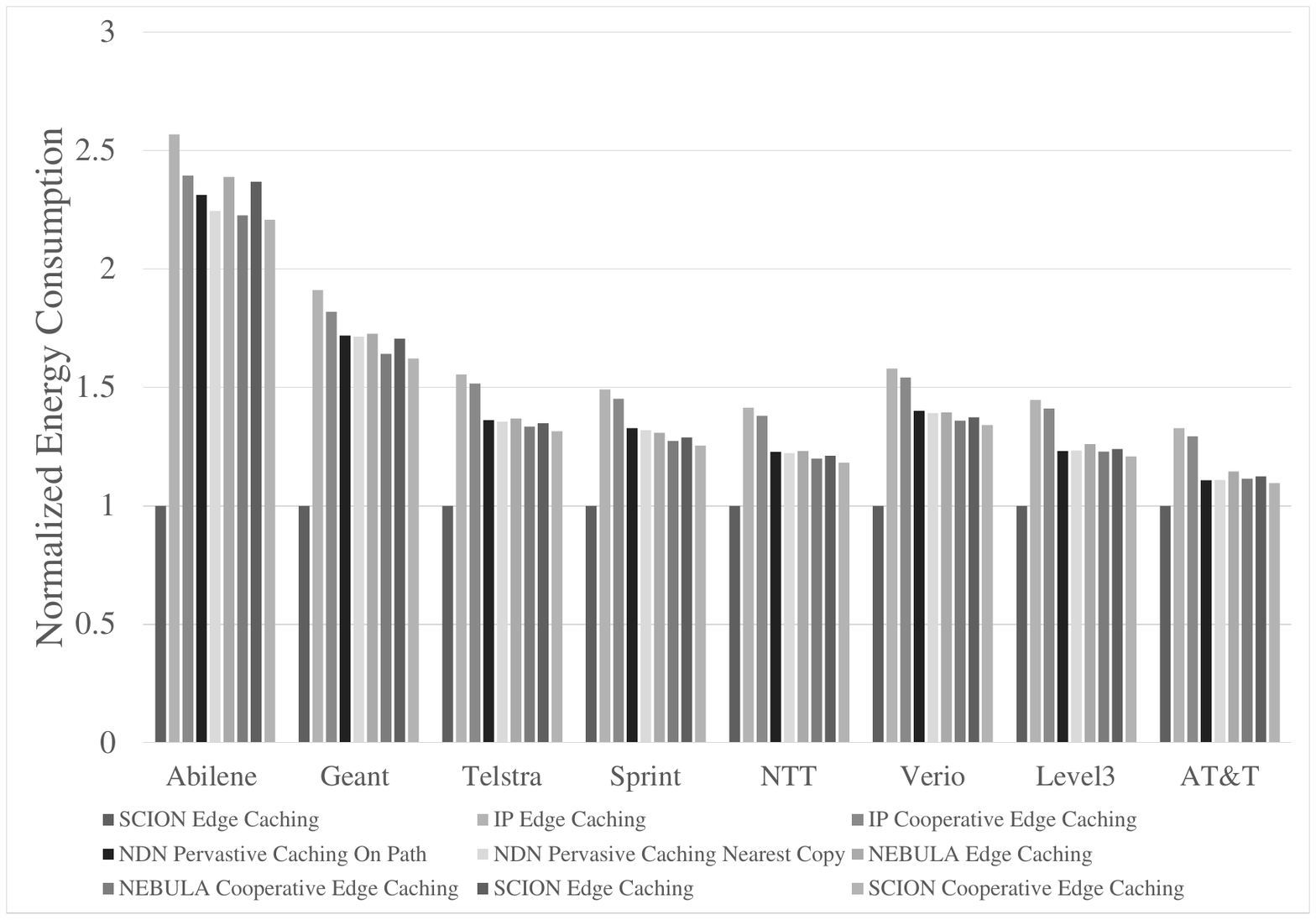}
		\label{fig:result_cache_strategy}
		\centering
	}
	\caption{(a) Power consumption of edge caching and pervasive caching with different cache budget ratios. All results are normalized by the power consumption of \scion with no cache. (b) Power consumption of edge caching and pervasive caching with different Zipf distribution parameters. All results are normalized by the \scion without cache. (c) Power consumption of edge caching and pervasive caching with different 
		cache discovery strategy. For edge caching, simple edge caching and cooperative 
		edge caching are considered. For pervasive caching, on-path cache discovery and 
		nearest-copy discovery are evaluated. All results are normalized by \scion 
		with edge caching.}
	\centering
\end{figure*}

\subsubsection{Cache budget ratio} 
For analyzing the influence of cache budget ratio on our observations, we 
vary the cache budget ratio from $c=0$, where no content cache is present 
in the network, to $c=1$, where all content in the Internet can be
cached in the network. Note that the current cache budget for major CDN
provider is around 5\% ($c=0.05$). Thus our analysis covers a wide range of 
values for the cache budget ratio parameters.

Figure~\ref{fig:result_cache_budget} graphs the power consumption of 
different networks with caching under different cache budget ratios. With 
increasing cache budget, the power consumption of different networks all 
increases. When the cache budget reaches 1, \ndn with pervasive caching, 
which consumes least amount of power among the candidates, still consumes
150\% more power than \scion with no caching.

On the other hand, the power consumption for \ndn with pervasive caching 
increases at a slower rate compared to \ip, \nebula, and \scion.
\ndn with pervasive caching consumes the same amount of 
power as SCION with edge caching when the cache budget is 0.76. \ndn with 
pervasive caching reduces up to 30\% the power budget in comparison to edge 
caching when the cache budget reaches 1. However, because the current cache 
budget for large CDN providers is around 0.05~\cite{fayazbakhsh2013less}, which 
is far smaller than 0.76, we expect our observation 3 still holds in the near 
future.

\subsubsection{Content Access Locality} 
For this analysis, we measure power consumption for different network 
architectures when varying Zipf distribution parameter $\alpha$. 
Larger $\alpha$ value indicates more locality in content 
access. With increasing locality in the content accesses, the both edge caching 
and pervasive caching will reduce energy footprints. 

In Figure~\ref{fig:result_zipf_dist}, we show that the energy footprints for 
\ip, \nebula, \scion with edge caching are still higher 
than that of \scion without cache even with high locality in the access 
pattern ($\alpha$=1.5 compared to the case when $\alpha$=0.99 in content 
accesses from US). On the other hand, for \ndn, its power efficiency will catch 
up with that of \scion without caching when $\alpha$ surpasses 1.1. With high 
locality ($\alpha$=1.5) \ndn{}'s power efficiency outperforms \scion without 
caching by 25\%. Therefore, our observation 4 will not hold when the Zipf 
distribution parameter is over 1.1. We note, however, that current $\alpha$ 
values are 0.99 for US, 0.92 for Europe, and 1.04 for Asia. Thus, for the time 
being, SCION appears to be the most power efficient architecture. 

\subsubsection{Cache Discovery Strategy}
For pervasive caching, we compare two cache discovery strategies: \emph{on-path 
	cache discovery}, in which only content cached by the on-path routers are
served, and \emph{nearest cache discovery}, in which the request of the content 
is redirected to the nearest router caching the content.  

For edge caching, we compare two cache discovery strategies: \emph{simple
	edge caching}, in which each caching server serves a content request only
from local caches, and \emph{cooperative cache discovery}, in which 
each caching server redirects content requests to the nearest servers that cache
the content. 

We realize that achieving nearest cache discovery and cooperative cache 
discovery require sophisticated mechanisms which in turn consume additional 
unaccounted power. However, by intentionally omitting the power 
consumed to distribute information about cached content to neighbour routers, 
we treat the nearest cache discovery and cooperative cache 
discovery as the optimal cases to characterize the boundaries of 
cache discovery strategy space. 

Figure~\ref{fig:result_cache_strategy} shows the power consumed by architectures 
with different caching strategies. We are interested in the differences in
power consumption that are caused by the various cache discovery strategies.
In general, the differences between different 
cache discovery strategies for both networks with edge caching and 
\ndn with pervasive caching lie 
within 2\%. Our observation 2, 3, and 4 hold with various cache strategies.

\subsubsection{Summary}
When varying the cache budget, all our 
observations hold. We see similar results when varying the cache discovery 
strategy. When varying the locality in access patterns, we note that certain
values for content access distribution weaken our observations. Specifically, 
with high locality, \ndn has the potential to be more power efficient than 
SCION. In addition, these high locality scenarios show a more positive effect 
on the benefits in power consumption due to pervasive caching. 

\begin{table}
	\begin{tabular}{p{5cm}|ccc}
		\hline
		Observation & CB & LAP & CDS \\\hline
		1. PCS consumes less power than lookups & \cmark & \cmark & \cmark \\\hline
		2. End-to-end consumes less power than caching & \cmark & \cmark & \cmark 
		\\\hline
		3. Pervasive caching offers marginal power consumption improvements & \cmark & 
		\xmark & \cmark \\\hline
		4. SCION consumes least amount of power & \cmark & \xmark & \cmark \\\hline
	\end{tabular}
	\centering
	\caption{Summary of the sensitivity analysis results. CB stands for cache 
		budget, LAP stands for locality in access pattern, CDS stands for cache 
		discovery strategy. A \cmark{} means that our observation remains the same when 
		the factor for the current column varies in the full range that we considered. 
		A \xmark{} means that our observation 
		remains the same when the factor for the current column is restricted in 
		a certain range.}\label{tab:sensitivityanalysis}
\end{table}

\section{Discussion}
\label{sec:discussion}\label{sec:disc}

In this paper, we concentrate on data-plane traffic power consumption in
content delivery scenarios. Bolla \etal~\cite{bolla2011energy} 
estimate that data-plane
traffic consumes 83\% of the total Internet power compared to 17\% consumed by 
control-plane traffic. Furthermore, content delivery applications constitute a 
majority of the Internet traffic. For example, Netflix and Facebook together 
account for 47\% of the downstream traffic today~\cite{Sandvine2013}. Our 
analysis assumes that the FIAs' control-plane still consumes a small proportion 
of the total power, and content delivery applications' traffic still constitutes
the majority of the Internet traffic.
Since implementation details regarding the FIAs' 
control-plane behaviors are not yet fully 
specified~\cite{Jacobson2009NDN, raychaudhuri2012mobilityfirst, 
Anderson2013NEBULA, Xin2011SCION}, in-depth analysis of control-plane 
power consumption behavior remains an open problem. 
In addition, the analysis presented herein does not capture power consumption behavior of 
real-time traffic, such as Skype communication. We defer the analysis of real-time traffic 
power consumption to future 
work, but we expect the results to be consistent with the observations in this paper.

\textbf{IP forwarding techniques}.
In our analysis of the fowarding-decision-making module, we have chosen TCAM 
as the underlying technology for \ip routers,
because it has been widely used by ASICs in commodity routers~\cite{ciscocatalyst6500}.
Admittedly, there are many alternative methods 
for searching routing-tables~\cite{gupta2001algorithms}. 
For example, Cisco builds ASICs in the CRS-1 router 
for making forwarding decisions based on a treebit map with reduced-latency 
DRAM (RLDRAM)~\cite{eatherton2004tree}. The detailed comparison of 
different forwarding hardware is out of scope for this paper.

\textbf{Content caches on routers}.
We assume \ndn routers use key-value stores to build content caches.
Key-value stores provide high query rates while minimizing power consumption. 
However, the key-value stores we analyze (HashCache and SILT), are both 
built for persistent storage. We expect that a key-value store built 
exclusively for caching can consume less power, and 
thus further reduces power consumption in \ndn when using a pervasive 
caching layer.

\section{Related Work}
\label{sec:related}

\textbf{Power consumption of routers, the Internet, and FIAs}. Power consumption 
of the Internet infrastructure has been well studied at various levels of 
granularity. Ye, Micheli, and Benini theoretically model the power consumption
of the switching fabric in routers by their electrical components, such as
capacitors~\cite{ye2002analysis}. Baliga et~al.~\cite{baliga2009energy}
and Tucker et~al.~\cite{tucker2008energy} model the power consumption of  
optical IP networks based based on the power 
consumption of individual heterogeneous routers, switches, etc. Lee 
\etal~\cite{Lee2011} estimate the power consumption of 
CCNs. Our paper adds to this body of work by analyzing both the power consumption of
FIA routers themselves, and large simulated FIA networks. Compared to 
previous work, our general power consumption model spans across two levels 
of granularity: 1) the power consumption of the computation 
and storage needed by different FIA router components; and 2) overall 
FIA network power consumption under various caching strategies and workloads. 

\textbf{FIA evaluation and metrics.} Previous work mainly focuses on 
the evaluation of content-centric networks (CCNs) or on the evaluation of 
specific CCN subsystems. Fricker 
\etal~\cite{fricker2012impact} evaluate the caching performance of 
CCNs under the influence of network traffic compositions. 
Muscariello, Carofiglio and Gallo~\cite{muscariello2011bandwidth} 
evaluate the performance of 
bandwidth and shared storage in CCN designs. 
Fayazbakhsh \etal~\cite{fayazbakhsh2013less} use network latency, network 
congestion and origin server load to compare different caching strategies in 
CCNs. Perino and Varvello~\cite{perino2011reality} use router 
throughput, monetary costs, and energy efficiency to estimate 
the feasibility of deployment of CCN routers. 
The main objective of our work is not limited to the exploration of
power consumption of each individual FIA. Rather, we concentrate on 
presenting an evaluation framework to explore the power 
implications of adopted FIA design principles. The results obtained herein 
can help guide designers toward power-efficient network designs.

\section{Conclusion and Future Work}
\label{sec:conclusion}\label{sec:concl}
In this paper we have modeled and compared the power consumption of future 
Internet architectures. We performed experiments at 
multiple levels of granularity ranging from per-bit power consumption of 
 router components, to network-wide power consumption under the use 
of different caching strategies. From our analysis, we were able to draw 
several observations: 1) the use of packet-carried state is more power-efficient than 
routing table lookups; 2) based on our workload assumptions, end-to-end 
communication consumes less power than using in-network caches; and 3) there is 
no substantial difference between energy footprints of networks with edge 
caching as compared to ones with pervasive caching. 

We propose power consumption as a general unified metric to optimize 
networks, as lower energy translates into smaller amounts of work performed. Thus,
 power minimization also optimizes the amount of equipment 
used, network performance, and environmental impact. We hope 
that our approach serves as a useful step toward making power analysis a common 
evaluation mechanism for network architectures.

 \section{Acknowledgments}
 The research leading to these results has received funding from the
European Research Council under the European Union's Seventh Framework
Programme (FP7/2007-2013) / ERC grant agreement 617605, and NSF under award number
CNS-1040801. We gratefully
acknowledge support from ETH Zurich and from the Zurich Information
Security and Privacy Center (ZISC). 
\Urlmuskip=0mu plus 1mu\relax
\bibliographystyle{plain}
\bibliography{0-string,bib}

\end{document}